\documentclass[12pt]{article} 
\let\section=\subsection     \let\subsection=\subsubsection

\usepackage{graphicx}
\usepackage{amsmath,amssymb,bm}

\newcommand{\vlk}{V_{\text{low}\,k}}
\newcommand{\be}{\begin{equation}}
\newcommand{\ee}{\end{equation}}
\newcommand{\bea}{\begin{eqnarray}}
\newcommand{\eea}{\end{eqnarray}}
\newcommand{\hf} {\frac{1}{2}}

\def\eq#1{(\ref{#1})}

\def\la{\langle}
\def\ra{\rangle}

\def\cD{{\mathcal{D}}}
\def\psid{\psi^\dagger}
\def\tr{{\mathrm{Tr}}}

\def\tG{\widetilde{G}}
\def\tGi{{\widetilde{G}}^{-1}}

\def\tGa{\widetilde{\Gamma}}
\def\mb#1{{\mathbf{#1}}}

\def\rgs{\rho_{\mathrm{gs}}}
\def\rgl{\rho_{\mathrm{gs},\lambda}}

\begin{document}
\begin{center}
{\large \bf Towards Density Functional Calculations} \\[2mm]
{\large \bf from Nuclear Forces} \\[5mm]
A.~Schwenk$^1$ and J.~Polonyi$^{2, 3}$ \\[5mm]
{\small \it  $^1$Department of Physics, The Ohio State University,
Columbus, OH 43210, USA \\
$^2$Institute for Theoretical Physics, Louis Pasteur University,
Strasbourg, France \\
$^3$Department of Atomic Physics, Lorand E\"otv\"os University,
Budapest, Hungary \\[8mm]}
\end{center}

\begin{abstract}
\noindent
We propose a method for microscopic calculations of nuclear ground-state 
properties in the framework of density functional theory (DFT).
We discuss how the density functional is equivalent to the effective 
action for the density, thereby establishing a constructive framework 
for DFT calculations from nuclear forces. The presented approach starts 
from non-interacting nucleons in a background potential (a simple
approximation for the mean field). The nuclear forces are then gradually 
turned on, while the background potential is removed. The evolution 
equation yields the ground-state energy and density of a system of 
interacting nucleons, including exchange-correlations beyond the RPA 
approximation. The method can start from non-local low momentum ($\vlk$) 
or chiral interactions.
\end{abstract}

\section{Introduction}

Microscopic approaches to large nuclei are inherently difficult, since
nucleons interact strongly and the number of possible configurations 
increases rapidly with the number of nucleons. As a consequence, for the
largest nuclei the computational method of choice has been mean field theory. 
A promising framework to extend and systematize mean-field calculations 
is density functional theory (DFT). DFT is the underlying theory in modern 
nuclear structure approaches to heavy nuclei, see e.g.,~[1-6].

In this talk, we discuss an approach complementary to parameterizing
the nuclear density functional. We present a DFT method to calculate 
nuclear ground-state properties from microscopic two-nucleon and 
few-nucleon forces. In a DFT framework, the dynamic degrees of freedom
are one-body densities, which simplify a microscopic description for larger
systems and are computationally feasible compared to solving for the
many-body wave function. In addition, DFT can start from the softer 
low momentum ($\vlk$) or chiral interactions~[7-13], which are non-local.
This is particularly important as the effective field theory approach 
offers at present the only consistent and practical counting scheme to 
construct three-nucleon forces. We start by showing that the effective 
action formalism provides a microscopic foundation for DFT. We focus
especially on how exchange-correlations are included. A direct
calculation of the effective action for non-interacting fermions
shows possible expansions of the density functional when interactions 
are included. We then present a functional RG-inspired method to 
calculate nuclear ground-state properties from microscopic interactions 
and discuss possible extensions.

\section{Effective action formalism and DFT}

We consider non-relativistic fermions in a time-independent 
background potential $V(\mb{x})$ (e.g., the ionic potential 
for electronic systems; for nuclei, $V=0$) interacting
through two-body forces $U(\mb{x},\mb{y})$. We note that 
three-body interactions can be easily included and for 
simplicity we also suppress isospin indices. The action $S$ 
consists of a one-body $S_{1}$ and an interacting two-body 
part $S_2$ (in Euclidean space-time with spin indices $\sigma$ and 
$\hbar = 1$)
\begin{align}
S[\psid,\psi] &= S_{1}[\psid,\psi]+S_{2}[\psid,\psi] = \sum_{\sigma} 
\int dx \: \psid_{\sigma}(x) \bigl[ \partial_t - \frac{1}{2m}\nabla^2_\mb{x} 
+ V_{\sigma}(\mb{x}) \bigr] \psi_{\sigma}(x) \nonumber \\
&+ \frac{1}{2} \sum_{\sigma_i,\sigma'_i} \int d\mb{x} \: d\mb{y} \:
dt \: \psid_{\sigma_1}(\mb{x},t) \: \psi_{\sigma_1'}(\mb{x},t) 
\: U_{\sigma_i,\sigma'_i}(\mb{x},\mb{y}) \: \psid_{\sigma_2}(\mb{y},t) 
\: \psi_{\sigma_2'}(\mb{y},t) .
\end{align}

The generating functional $W[K]$ for connected density correlation
functions is given by a path integral over all configurations\footnote{For 
simplicity, we introduce the 
short-hand notation $\psid \cdot \psi \equiv \sum_{\sigma} 
\int dx \: \psid_{\sigma}(x) \: \psi_{\sigma}(x)$ or
$(\psid \psi) \cdot M \equiv \sum_{\sigma,\sigma'} \int dx \: 
\psid_{\sigma}(x) \: \psi_{\sigma'}(x) \: M_{\sigma,\sigma'}(x)$.
This allows to rewrite the action in compact form $S_{1}[\psid,\psi]
\equiv \psid \cdot G^{-1} \cdot \psi$ and $S_{2}[\psid,\psi] \equiv
\frac{1}{2} \: (\psid\psi) \cdot U \cdot (\psid\psi)$, where the
free propagator $G$ includes the background potential $V$.}
\be
e^{W[K]} = \int\cD[\psid]\cD[\psi] \: e^{-S[\psid,\psi]+(\psid\psi) \cdot K} ,
\label{genfunc}
\ee
where we have introduced spin-dependent sources $K_{\sigma,\sigma'}(x)$ 
coupled to the possible density operators $\psid_{\sigma}(x) \: 
\psi_{\sigma'}(x)$. The strategy for introducing the source term 
is (loosely) analogous to the calculation of the magnetization of 
spins on a lattice. For spin systems, one introduces a
magnetic source $H$ coupled to the total spin of the system, $H \, 
\sum_i S_i$. The variation of the source $H$ then probes all different 
configurations and allows one to determine the spin configuration, 
i.e., the magnetization, which minimizes the free energy. For ground-state 
properties of many-body system, the source is coupled to the 
density operator and variations probe different density profiles. Thus,
by varying the generating functional $W[K]$ with respect to the source
$K$ one obtains the density $\rho_{\sigma,\sigma'}(x)$ in the presence 
of the source (and the ground-state density for $K=0$)
\be
\rho_{\sigma,\sigma'}(x) \equiv \la \psid_{\sigma}(x) \: 
\psi_{\sigma'}(x) \ra_{K} = \frac{\delta W[K]]}{\delta
K_{\sigma,\sigma'}(x)} .
\label{rhodef}
\ee
Assuming that Eq.~\eq{rhodef} is invertible, it can be used to express 
the source in terms of the density $K[\rho]$, and one can perform
a functional Legendre transformation from $W[K]$ to the effective 
action $\Gamma[\rho]$ given by
\be
\Gamma[\rho] = -W[K] + K \cdot \rho .
\label{legtransf}
\ee
The effective action has the advantageous property that it is minimal
at the physical (zero source) ground-state density $\rho=\rgs$
\be
\frac{\delta \Gamma[\rho]}{\delta\rho} \biggr|_{\rho=\rgs} = 0 .
\label{mincond}
\ee
Moreover, the effective action at the minimum yields the
ground-state energy of the system $E_{\text{gs}}$. This follows from
the spectral representation of the path integral $e^{W} = \sum_n 
e^{-\beta E_n}$ and for $K=0$ one has in the zero temperature limit
\be
E_{\text{gs}} = E[\rgs] = \lim\limits_{\beta \to \infty} \frac{1}{\beta}
\, \Gamma[\rgs] .
\label{energyfunc}
\ee
Summarizing, the effective action is a functional of the density which is
minimal at the ground-state density and reproduces the energy. Therefore,
the effective action for the density is equivalent to the energy density 
functional in the sense of Hohenberg-Kohn~\cite{HK} (more precisely for 
time-independent sources). This has been pointed out in the literature, 
see e.g., recently~[15-18].\footnote{Two more comments are
in order. First, a chemical potential corresponds to a constant source,
which requires some attention in the calculation of the effective action
and leads to a minimum condition $\delta \Gamma[\rho] / \delta \rho 
|_{\rho=\rgs} = \mu$ for homogeneous systems, since the source and the 
chemical potential cannot be distinguished. Second, while the densities
are to be expanded in a set of orbitals, $\rgs(\mb{x}) = \sum_n c_n 
|\phi_n(\mb{x})|^2$, density fluctuations such as $\rho(x)-\rgs(x)$ must 
be expanded in a set of basis functions whose integral vanishes.}
One advantage of the effective action formalism is that it presents a 
constructive framework for microscopic calculations of the density 
functional. For completeness, we further prove the Hohenberg-Kohn 
universality (i.e., the trivial $V$ dependence) of the density functional. 
This simply follows by noticing that the one-body potential $V$ enters 
in the many-body dynamics in the same way as the external source $K$. 
Therefore, the dependence of the generating functional $W[K]$ on $V$ 
is linear, $W[K] = W_{V=0}[K-V]$ and the Legendre transformation gives 
\be
\Gamma[\rho] = -W[K] + K \cdot \rho = -W_{V=0}[K-V] + (K-V+V) \cdot \rho 
= \Gamma_{V=0}[\rho] + V \cdot \rho ,
\ee 
where $\widetilde{K} = K-V$ is taken as source for the Legendre
transformation.

\section{Non-interacting fermions}

After discussing the microscopic basis of DFT using effective action
techniques, we consider the non-interacting case. This provides insight 
into physical expansions of the effective action and how exchange-correlations 
will be included. In the non-interacting case, the generating functional 
is a Gaussian integral over the fermion fields and one has
\be
W_{\text{free}}[K] = \tr\log[\,G^{-1}-K\,] = \tr\log G^{-1}
- \sum\limits_{n=1}^\infty \frac{1}{n} \: \tr \bigl[\left(G \cdot K
\right)^n \bigr] .
\ee
The density in the presence of the source is given by 
$\delta W_{\text{free}}[K] / \delta K_{\sigma,\sigma'}(x)$,
\be
\rho_{\sigma,\sigma'}(x)
= - \, G_{\sigma',\sigma}(x,x) - \sum\limits_{n=0}^\infty 
G_{\sigma',..}(x,..) \cdot K \cdot (G \cdot K)^n \cdot 
G_{..,\sigma}(..,x) ,
\label{rhoK}
\ee
where the source is diagonal in the space-time index $x$ and
summed over indices are indicated by $..\:$. For $K=0$, we
have the familiar expression for the ground-state density,
$\rho_{\text{gs};\sigma,\sigma'}(x) = - \, G_{\sigma',\sigma}(x,x)$.
A second variation of the generating functional yields the
particle-hole propagator $\tG$ (or density-density correlator). From
Eq.~(\ref{rhoK}), we find for vanishing source in the ground state
\begin{align}
\tG_{X,Y} &\equiv \tG_{(\sigma_1,\sigma_1',x),(\sigma_2,\sigma_2',y)}
= \frac{\delta^2 W_{\text{free}}}{\delta K_{\sigma_2,\sigma_2'}(y)
\, \delta K_{\sigma_1,\sigma_1'}(x)} \biggr|_{K=0} \nonumber \\[1mm]
&= \la \psid_{\sigma_1}(x) \: \psi_{\sigma_1'}(x) \: 
\psid_{\sigma_2}(y) \: \psi_{\sigma_2'}(y) \ra
= - \, G_{\sigma_1',\sigma_2}(x,y) \: G_{\sigma_2',\sigma_1}(y,x) ,
\end{align}
where we have introduced a combined index $X=(\sigma_1,\sigma_1',x)$. It
is important to realize that the second functional derivatives of the
generating functional and the effective action are related (as for
thermodynamic potentials conjugate by a Legendre transformation). Thus, one
has
\be
\left( \frac{\delta^2 \Gamma[\rho]}{\delta \rho_X \, \delta \rho_Y} 
\right)^{-1} = \frac{\delta^2 W[K]}{\delta K_X \, \delta K_Y} .
\label{secondder}
\ee
Eq.~(\ref{secondder}) relates the dressed particle-hole propagator
(for the interacting system) to the curvature of the effective action 
in the ground state. We will later see that all exchange-correlations
are built up in this way.

For the Legendre transformation, we need the inverse expression to 
Eq.~(\ref{rhoK}), i.e., the source as a function of density. By 
rewriting Eq.~\eq{rhoK} in compact form
\be
\rho_X = \rho_{\text{gs};X} + \tG_{X,..} \cdot K_{..} + {\mathcal O}(K^2) ,
\ee
it is clear that the source $K$, and thus the generating functional $W[K]$ 
can be expanded in a power series in $\rho-\rgs$ around the ground-state 
density. For the source, one similarly has
\be
K_X =  - \tGi_{X,..} \cdot (\rho-\rgs)_{..} + {\mathcal O}\bigl(
(\rho-\rgs)^2 \bigr) .
\ee

We can now perform the Legendre transformation order-by-order, where the 
effective action has the same expansion around the ground-state
\be
\Gamma_{\text{free}}[\rho] = \Gamma_{\text{free}}^{(0)}
+ \sum_{n=2}^\infty \: \int_{X_1,\ldots,X_n} \frac{1}{n!}
\: \Gamma^{(n)}_{\text{free};X_1,\ldots,X_n} \, (\rho-\rgs)_{X_1} \cdots
\, (\rho-\rgl)_{X_n} ,
\ee
with ground-state energy $E_{\text{gs}} = \lim\limits_{\beta \to \infty} 
\Gamma_{\text{free}}^{(0)} / \beta$. Moreover, by keeping the 
$X$-dependence of the expansion coefficients, no local density approximation
has been made. A straightforward calculation of the expansion coefficients
yields
\begin{align}
\Gamma_{\text{free}}^{(0)} &= -\tr\log G^{-1} , \\[1mm]
\Gamma^{(2)}_{\text{free};X,Y} &= \tGi_{X,Y} , \\[1mm]
\Gamma^{(3)}_{\text{free};X_1,X_2,X_3} &= 2 \,
\int_{Y_1,Y_2,Y_3} S^{(3)}_{Y_1,Y_2,Y_3} \, \tGi_{Y_1,X_1} \,
\tGi_{Y_2,X_2} \, \tGi_{Y_3,X_3} ,
\end{align}
where $S^{(3)}$ denotes the totally symmetric $3$-propagator ring
\be
S^{(3)}_{X_1,X_2,X_3} = \frac{1}{3!} \bigl( G_{\sigma_1',\sigma_2}(x_1,x_2)
\, G_{\sigma_2',\sigma_3}(x_2,x_3) \, G_{\sigma_3',\sigma_1}(x_3,x_1)
+ \text{permutations} \bigr) .
\label{3propring}
\ee
Higher-order expansion coefficients have a similar structure as $\Gamma^{(3)}$.
With the expansion around the ground state, we next present a method 
that includes the interactions among nucleons in a non-perturbative manner. 
It is reminiscent of the quasiparticle picture and the adiabatic turning 
on of the interaction.

\section{Evolution equation for the density functional}

\begin{figure}[h!]
\begin{center}
\includegraphics[scale=0.5,clip=]{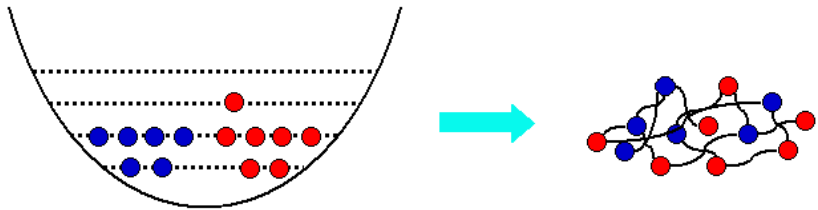}
\end{center}
\caption{A graphical illustration of the initial condition of the evolution
equation (left figure) and the final, physical system of nucleons
(right figure). The evolution equation interpolates between the two cases.}
\label{dftrg}
\end{figure}

We are interested in solving for the ground-state properties of the
physical system, where the nucleons interact by means of the microscopic 
interactions, without the presence of a background potential. To this
end, we introduce a control parameter $\lambda$, with $0 \leqslant 
\lambda \leqslant 1$, by replacing
\be
V \rightarrow (1-\lambda) \, V_\lambda \hspace{1cm} 
U \rightarrow \lambda \, U ,
\ee
where we allow the background potential $V$ to depend on $\lambda$,
as long as the corresponding eigenfunctions are localized and physically 
reasonable. At $\lambda=0$, when the two- and higher-body interactions 
are absent, the system is trivially bound by employing an attractive
background potential (a simple guess for the mean field). The nucleons 
fill up the lowest-lying states to the Fermi energy. (Generally, the 
initial background potential should be chosen to improve the convergence 
of the presented algorithm.) For $\lambda=1$, we have the physical 
system, where the background potential is switched off. We use the control
parameter to evolve a simple shell-model configuration to the
fully-interacting system including many configurations. This is 
depicted in Fig.~\ref{dftrg}. We note that such an RG-inspired method 
does not require the coupling strength to be small. The smallness of 
the interaction in perturbation theory is taken over by the differential 
change, $\delta \lambda$ serving as a ``small parameter''.

As the control parameter is gradually increased from $\lambda=0$ to 
$\lambda=1$, the density functional follows an evolution equation
given by~\cite{DFTRG}
\be
\partial_\lambda \Gamma_\lambda[\rho]=
-\partial_\lambda W_\lambda[K] = \partial_\lambda [(1-\lambda) \, 
V_\lambda] \cdot \rho + \frac{1}{2} \, \rho\cdot U \cdot\rho +
\frac{1}{2} \, \tr \left[ U \cdot 
\left( \frac{\delta^2\Gamma_\lambda[\rho]}{\delta\rho \, \delta\rho}
\right)^{-1} \right] ,
\label{rge}
\ee
where the derivation is similar to~\cite{Janos}. Since Eq.~\eq{rge} 
is exact, the solution for general density configurations is 
cumbersome and is not directly related to physical quantities. Therefore, 
we propose to solve the evolution equation only in the vicinity of 
the physical point, i.e., expanded around the evolving ground-state 
density $\rgl$. Before we discuss further details of the approach, 
we separate off the background and Hartree contributions to
the density functional, since their density dependence is known
exactly. Thus, we introduce the kinetic and exchange-correlation
part $\tGa_\lambda[\rho]$ of the density functional through
\be
\Gamma_\lambda[\rho] = (1-\lambda) V_\lambda \cdot \rho + 
\frac{\lambda}{2} \, \rho \cdot U \cdot \rho + \tGa_\lambda[\rho] ,
\label{treeoff}
\ee
which leads to a simplified evolution equation for $\tGa_\lambda[\rho]$
\be
\partial_\lambda\tGa_\lambda[\rho] = \frac{1}{2} \, \tr \left[ U \cdot \left(
\frac{\delta^2 \tGa_\lambda[\rho]}{\delta\rho \, \delta\rho} + \lambda \, U 
\right)^{-1} \right] .
\label{rge2}
\ee

As the ground-state density changes under the RG, we expand the
effective action around the current ground state $\rgl$ of the
evolving system. This is the underlying theme and efficacy of 
our RG-inspired approach. As in the non-interacting case, we have
\be
\tGa_\lambda[\rho] = \tGa_\lambda^{(0)}
+ \sum_{n=1}^{N_\Gamma} \: \int_{X_1,\ldots,X_n} \frac{1}{n!}
\: \tGa^{(n)}_{\lambda;X_1,\ldots,X_n} \cdot (\rho-\rgl)_{X_1} \cdots
(\rho-\rgl)_{X_n} .
\label{expansion}
\ee
We make the approximation and truncate the expansion for $\tGa_\lambda$ 
at a given order $N_\Gamma$ in density fluctuations. 
Since one expects five- and higher-body correlations 
to be relatively small, due to the Pauli principle, a truncation at
$N_\Gamma=3$ or $N_\Gamma=4$ seems a reasonable choice. Obviously,
the convergence of such an expansion must be checked by including 
higher-order contributions. (We note that, as for the Fermi liquid RG
approach used in~\cite{RGnm1,RGnm2}, the flow equation generates upon 
integration an infinite set of diagrams in the underlying two-body 
interaction $U$.)

\begin{figure}
\begin{center}
\begin{picture}(390,54) 
\put(16,1){\includegraphics[scale=0.49,clip=]{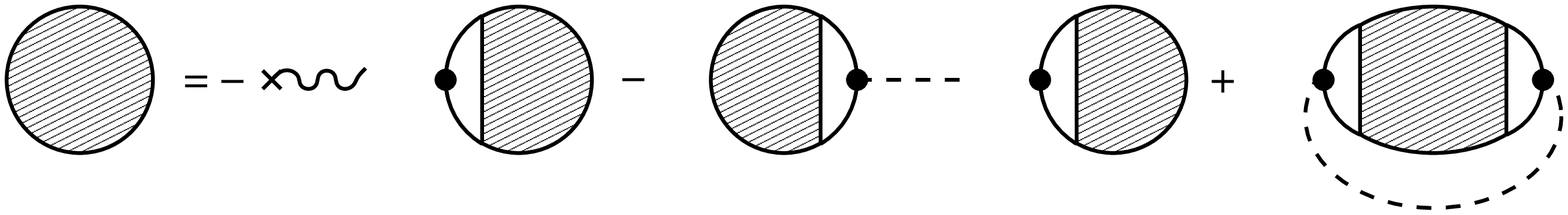}}
\put(1,29){{\large $\partial_\lambda$}}
\put(106,29){{\large $\partial_\lambda$}}
\put(176,29){{\large $\lambda$}}
\put(250,29){{\large $\partial_\lambda$}}
\put(320,29){{\Large $\hf$}}
\end{picture}
\end{center}
\caption{Graphical representation of the RG-like equation for 
the ground-state energy $E_{\text{kxc}}$. $E_{\text{kxc}}$, the ground-state 
density $\rgl$ and the density-density correlator $\tG_{\lambda}$ 
are denoted by a blob with $n = 0, 1, 2$ dots respectively, which 
label the index structure. $(1-\lambda) V_ \lambda$ is represented 
by a wiggly line and the two-body interaction $U$ by a dashed line. 
Open dots further label open indices, and filled dots represent 
indices which are traced over.}
\label{Egsevol}
\end{figure}

The truncation of the evolution equation at three-density correlations,
$N_\Gamma=3$, leads to coupled RG-like equations for the expansion 
coefficients, i.e., for the kinetic and 
exchange-correlation (kxc) energy $E_{\text{kxc}}$, the evolving 
ground-state density, the dressed particle-hole propagator 
$\tG_\lambda = \bigl( \tGa^{(2)}_\lambda + \lambda U \bigr)^{-1}$ 
and the three-density correlator $\tGa^{(3)}_\lambda$. To provide 
further insight, we focus on the flow equation for the ground-state 
energy and density, given by~\cite{DFTRG}
\be
\beta \, \partial_\lambda E_{\text{kxc}}[\rgl] \equiv
\partial_\lambda \tGa^{(0)}_\lambda = - (1-\lambda) V_\lambda \cdot 
\partial_\lambda\rgl - \lambda \, \rgl \cdot U \cdot \partial_\lambda\rgl
+ \hf \, \tr \left[ U\cdot \tG_\lambda \right] ,
\label{expc0}
\ee
Here, the contributions are due to the change in the ground-state density 
(first and second terms in Eq.~\eq{expc0}) and due to the dressed 
particle-hole propagator $\tG_\lambda$ (last term). The flow equation
is shown diagrammatically in Fig.~\ref{Egsevol}. The last term 
in Eq.~\eq{expc0} explicitly reads $\tr \left[ U\cdot \tG_\lambda 
\right] \equiv \int_{Y_1,Y_2} U_{Y_1,Y_2} \, \tG_{\lambda;Y_2,Y_1}$, 
and thus to lowest order is the Fock contribution to the energy. 
This establishes that all exchange-correlations are built in through 
the dressed particle-hole propagator $\tG_\lambda$. (Obviously, such 
a theory with densities as variables is not mean-field theory.) 
Moreover, we note that one recovers the standard RPA approximation
when only $\tG_\lambda$ is maintained, i.e., $N_\Gamma=2$ for the
effective action. The flow equation for the ground-state density 
is given by
\begin{align}
\partial_\lambda \rho_{\mathrm{gs},\lambda;X} &=
- \partial_\lambda[(1-\lambda) \, V_ \lambda] \cdot \tG_{\lambda;X}
- \rgl \cdot U \cdot \tG_{\lambda;X} \nonumber \\[1mm]
&+ \hf \, \tr \left[ U\cdot \tG_\lambda
\cdot \left( \tGa^{(3)}_\lambda \cdot \tG_{\lambda;X} \right) \cdot
\tG_\lambda \right] ,
\label{expcgr}
\end{align}
where the expansion coefficients are totally symmetric and the indices not
written explicitly are traced over. In the first iteration of the
flow equation, $\tGa^{(3)}$ is given by the $3$-propagator ring with
three inverse density-density correlators, which cancel with the three
$\tG$ in the last term of Eq.~\eq{expcgr}. The remaining exchange 
contributions from the $3$-propagaptor ring $\tr \left[ U \cdot 
S^{(3)}_{\lambda;X} \right]$ will lead to a quasiparticle-like distribution 
over the initially unoccupied orbitals.

\section{Conclusions and extensions}

In summary, the effective action formalism is a constructive framework 
for the density functional. It allows for calculations of ground-state 
properties starting from microscopic two- and three-nucleon 
interactions, which can be non-local such as the soft low momentum 
($\vlk$) or chiral interactions~[7-13]. 
The main advantage of DFT lies in using densities
as degrees of freedom, as compared to many-body wave functions. As
a consequence, microscopic DFT approaches should scale well to larger
nuclei. There is some insight from electronic systems, where DFT (see
e.g.,~\cite{Dreizler}) and 
the coupled cluster method give overall good results for ground-state 
properties in larger systems.

We believe that the presented DFT approach may be a step towards the
derivation of nuclear structure of heavier systems
from nuclear forces. We are currently working on the feasibility
of this approach for strongly-interacting one-dimensional models~\cite{DFTRG}.
Systematic, microscopic approaches are particularly important for 
extrapolations to neutron/proton-rich systems, e.g., due to the
constraints on the isospin dependence from nuclear forces or to maintain
a complete basis of induced non-central interactions~\cite{tensor}.

Extensions of this method are under investigation. A projection on 
vanishing center-of-mass kinetic energy can be implemented by introducing
sources for the center-of-mass motion. The harmonic oscillator frequency 
of the background potential can be adapted under the evolution to improve 
the convergence (reflecting the variational character of the single-particle
basis). Explicit pairing correlations can be introduced by coupling
a source to the off-diagonal densities $\psi_{\sigma}(x) \: \psi_{\sigma'}(x)$
(and the Hermitian conjugate). The additional minimum condition of the
effective action is then equivalent to the BCS gap equation.

\subsection*{Acknowledgments}

It is a pleasure to thank Dick Furnstahl for many useful discussions.
The work of AS is supported by the NSF under Grant No. PHY-0098645.

\end{document}